\documentclass[aps,twocolumn,showpacs,preprintnumbers]{revtex4}
\usepackage{graphicx}
\begin{document}
\preprint{accepted by Phys. Rev. E}
\title{Jamming transition in a highly dense granular system under vertical vibration}
\author{Kipom Kim}
\email{kpkim@pusan.ac.kr}
\author{Jong Kyun Moon}
\author{Jong Jin Park}
\author{Hyung Kook Kim}
\author{Hyuk Kyu Pak}
\email{hkpak@pusan.ac.kr} \affiliation{Department of Physics, Pusan
National University, Busan, 609-735, Korea}
\date{January 12, 2005}
\begin{abstract}
The dynamics of the jamming transition in a three-dimensional
granular system under vertical vibration is studied using
diffusing-wave spectroscopy. When the maximum acceleration of the
external vibration is large, the granular system behaves like a
fluid, with the dynamic correlation function $G(t)$ relaxing
rapidly. As the acceleration of vibration approaches the
gravitational acceleration $g$, the relaxation of $G(t)$ slows down
dramatically, and eventually stops. Thus the system undergoes a
phase transition and behaves like a solid. Near the transition
point, we find that the structural relaxation shows a stretched
exponential behavior. This behavior is analogous to the behavior of
supercooled liquids close to the glass transition.
\end{abstract}
 \maketitle

\section{Introduction}
Granular materials are the particle systems in which the size of
particle is large and the effect of thermal agitation is
negligible\cite{Review}. Recently there has been much interest in
the physics of noncohesive granular materials lying on a vertically
vibrating surface. When the vibration intensity is large, the
granular systems show the properties of fluids, such as
convection\cite{Convection1,Convection2}, heaping\cite{Heaping,Gas},
traveling surface waves\cite{Pak}, pattern formation\cite{Swinney,
Kim}, and size segregation\cite{Segre}. When the vibration intensity
is small, disordered granular materials become jammed, behaving like
a system with infinite viscosity. After Liu and Nagel proposed an
idea unifying the glass transition and the jamming
behavior\cite{Liu}, the behavior of the jamming transition has been
studied
extensively\cite{Weitz,OHern,Coniglio,Nicodemi,Bideau,DAnna1,Bonn,Powders}.
In this model, the inverse density of the system $\rho^{-1}$,
temperature $T$, and stress $\sigma$ form the axes of a
three-dimensional phase diagram, with the jammed state in the inner
octant and the unjammed state outside. In athermal macroscopic
systems like granular materials, thermal temperature does not play
any important role. Instead, it is an effective temperature that
relates the random motion  of the particles.  When jammed by
lowering the effective temperature, the system is caught in a small
region of phase space with no possibility of escape. For thermal
systems, if the molecules are bulky and of irregular shape, or if
the liquid is cooled too rapidly for the crystalline structure to
form, at low temperature it vitrifies into a rigid phase that
retains the disordered molecular arrangements of the liquid,
creating a glass state. If the idea of the jamming phase diagram is
correct, one can apply many ideas of the glass transition to explain
the jamming behavior in athermal systems.

In order to study the dynamics of granular systems, one might want
to track the motion of the individual particles. However, due to the
opacity of the granular systems, most experimental work has been
limited to the study of the external features of the granular flow
or the motion of tracer particles. Recently, noninvasive
experimental techniques using magnetic resonance imaging (MRI),
positron emission particle tracking (PEPT), and diffusing-wave
spectroscopy (DWS) have overcome the problem of opacity in the
three-dimensional granular system and have made observation of
internal features of the flow
possible\cite{Convection1,Pept,Durian,You}. Since the spatiotemporal
scale of the particle fluctuations is much smaller than the
resolution of MRI and PEPT, only DWS can provide a statistical
description of a highly dense system of small particles with
adequate resolution\cite{DWS}. To briefly explain the DWS technique,
photons are scattered consecutively by many particles in a highly
dense medium. This diffusive nature of light in a strongly scattered
medium results in a scattered light intensity that fluctuates with
time. Thus the temporal decay of the light intensity autocorrelation
function is used to study very small relative motions of the
scatterers in the medium. The DWS technique has been successfully
applied to study the motion of granular particles in a channel flow,
a gas-fluidized bed, an avalanche flow, and a vibro-fluidized
bed\cite{Durian,You}.

In this paper, the fluidization and jamming process of a thick and
highly dense three-dimensional vibro-fluidized granular bed is
studied using DWS. Exploring the temperature axis of the jamming
phase diagram\cite{Liu}, we compare our experimental result with
theoretical concepts developed in the study of supercooled liquids
close to the glass transition. Our result provides strong evidence
of the analogy between the dynamics of granular materials and the
behavior of supercooled liquids close to the glass transition.

\section{Experimental Setup}

\begin{figure}
\center{\includegraphics[width=7.5cm]{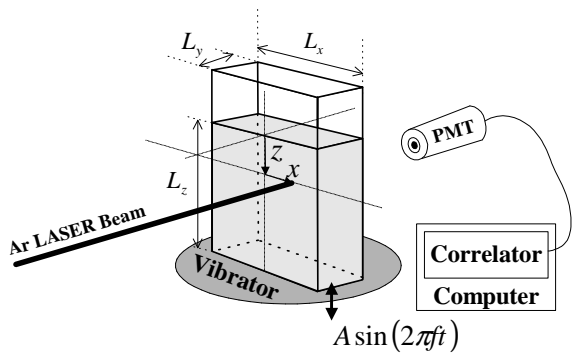}} \caption{Schematic
diagram of experiment setup. \label{fig:setup}}
\end{figure}

Figure 1 is a simple schematic diagram of the experimental setup. A
180 mm high, 90 mm wide ($L_x$), and 10 mm thick ($L_y$) rectangular
glass vessel is filled with glass beads of diameter $270 \pm 20
\mu$m at the total depth of $L_z=100$mm. An electromagnetic shaker
vibrates the vessel vertically with the form  $A~{\rm sin}\left(2\pi
f t\right)$, where $A$ is the amplitude and $f$ is the frequency of
the vibration. One can construct a dimensionless acceleration
amplitude, whose maximum value is
\begin{equation}
\Gamma = 4 \pi^2  f^2 A/g ,
\end{equation}
where $g$ is the gravitational acceleration\cite{Heaping}. The
vessel is much heavier than the total mass of the glass beads so
that the collisions of the vessel with the granular medium do not
disturb its vertical vibration. The air inside the vessel is
evacuated below the pressure of 0.1Torr, where the volumetric effect
of the gas can be neglected\cite{Gas}. The acceleration amplitude
$\Gamma$ and the vibration frequency  $f$ are monitored by the
accelerometer which is positioned on the vessel. All the data are
taken in a steady state condition where the system is well
compacted\cite{Bideau}. For the DWS measurements, an Ar laser beam
with the wavelength of 488nm is directed on the wide side of the
vessel, and the transmitted light intensity is detected by a
photomultiplier tube (PMT). The center of scattering volume is
located at the position ($x,z$), where $x$ is the horizontal
distance from the central-vertical axis of the vessel and $z$ is the
vertical distance below the free surface. The intensity output $I$
of the PMT is fed to a computer-controlled digital correlator
(BI9000AT; Brookhaven Inst.). The intensity autocorrelation
function, which is calculated by the digital correlator, is defined
as
\begin{equation}
g_2 (t) = <I(t) I^* (0)>  / <I(0)>^2 = 1 + \kappa G (t) ,
\end{equation}
where $t$ is a delay time, $\kappa$ is a factor determined by the
optical geometry of the experiment\cite{Chu}, and the bracket
indicates a time average. In a highly dense granular medium,
incident photons are multiply scattered by the glass beads. When the
beads in the medium move, the intensity $I$ measured at the detector
fluctuates with time. Thus the intensity autocorrelation function
has the information of the movement of the beads in the medium. When
$I$ is periodic, there are echoes in the correlation function at the
integral multiples of the oscillation period\cite{Echo}. In this
experiment, the position of the incident laser beam is fixed and the
scattering medium vibrates with the form of $A~{\rm sin}\left(2\pi f
t\right)$. Therefore when all the particles in the scattering volume
exhibit perfectly reversible periodic motion, the correlation
function will return to its initial value 1[=$G(0)$] at each
multiple of the oscillation period. When only a fraction of the
scatterers undergoes a reversible motion in the scattering volume,
the height of the echoes will decay in time. Lastly, when all the
scatterers exhibit an irreversible motion, there is no echo in the
correlation function. Therefore the degree of fluidization can be
characterized by the decay time of the height of the echoes in
$G(t)$.

\section{Results and Discussion}

\subsection{Fluidization point}

\begin{figure}
\center{\includegraphics[width=7.5cm]{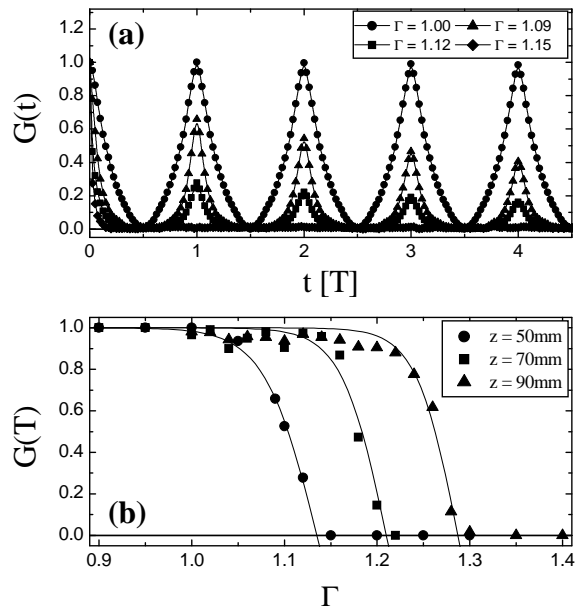}} \caption{ (a) The
correlation functions measured using DWS at various dimensionless
acceleration $\Gamma$ in the case of the position ($x=0$mm,
$z=50$mm) and $f=50$Hz. The unit of time in this figure is the
vibration period $T$. (b) The height of the first echo in the
correlation function at three different depths ($z$) in the case of
$x=0$mm and $f=50$Hz. \label{fig:correlation}}
\end{figure}

Figure 2(a) shows the correlation function at various dimensionless
accelerations at the position ($x=0$mm, $z=50$mm) in the case of the
vibration frequency of $f=50$Hz and the granular bed of $L_z=100$mm.
When the maximum acceleration is smaller than the gravitational
acceleration, $\Gamma<1$, the correlation function oscillates with
echoes at every multiple of the period $T$($\equiv f^{-1}$). Here,
the height of each echo does not change, that is $G_{echo}(t)=1$.
Since the photons meet the same scattering volume at every vibration
period, $G_{echo}(t)=1$ implies that there is no relative movement
of the particles in the scattering volume. This suggests that the
granular system is solidlike at  $\Gamma<1$, where the viscosity of
the system appears to diverge. As the dimensionless acceleration is
increased above $\Gamma = 1$, however, the height of the echoes in
the correlation function relax in time. At a certain critical
dimensionless acceleration $\Gamma_m$, the height of the first echo
disappears completely, [$G(T)=0$]. For $\Gamma > \Gamma_m$, $G(t)$
decays very rapidly without any echoes. This implies that at $\Gamma
> \Gamma_m$, the particle positions become completely randomized by
the external vibrations and the statistical description of the
internal motion becomes important. In this sense, $\Gamma_m$ is the
critical dimensionless acceleration where the scattering volume
becomes fluidized. This fast decay in $G(t)$ at $\Gamma > \Gamma_m$
contains information about the short-time dynamics of the granular
particles\cite{You,DAnna2}. Figure 2(b) shows the height of the
first echo $G (T)$ as a function of $\Gamma$ at different vertical
positions $z$. For a given position inside the vessel, the height of
the first echo decreases monotonically with $\Gamma$. Thus the
fraction of the particles which undergo an irreversible motion
increases with $\Gamma$. To determine $\Gamma_m$, the data points
are fitted with the function, $\tanh [\alpha (\Gamma - \Gamma_m)]$.

\begin{figure}
\center{\includegraphics[width=7.5cm, height=10cm]{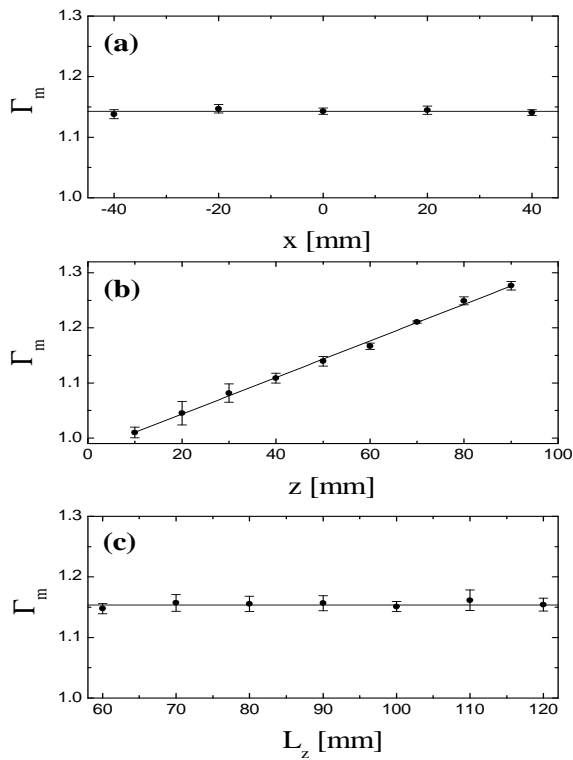}}
\caption{ The critical dimensionless acceleration $\Gamma_m$ at
various position. Each data point is the averaged value from many
independent measurements. (a) $\Gamma_m$ at various horizontal
position ($x$) in the case of $z=50$mm and $f=50$Hz. (b) $\Gamma_m$
at various vertical positions ($z$) in the case of $x=0$mm and
$f=50$Hz. (c) $\Gamma_m$ at various total depths $L_z$ in the case
of the position ($x=0$mm, $z=50$mm) and $f=50$Hz.
\label{fig:position}}
\end{figure}

Figure 3(a) shows the critical dimensionless acceleration $\Gamma_m$
at various horizontal positions $x$ with $z$=50mm and $f=50$Hz. Note
that $\Gamma_m$ is independent of the horizontal position. Figure
3(b) plots $\Gamma_m$ as a function of the vertical position $z$
with $x=0$mm in the case of $f=50$Hz and $L_z=100$mm. The critical
value of the dimensionless acceleration increases linearly with the
distance from the free surface, except near the free surface of the
granular system. The solid line in this figure is a straight line
with a slope of 0.0034. The result that $\Gamma_m$ is nearly $1$
near the free surface of the granular system agrees with many
previous experimental studies\cite{Convection2,Heaping,Gas,Pak}.
Figure 3(c) shows $\Gamma_m$ at various total depths $L_z$ in the
case of the fixed position ($x=0$mm, $z=50$mm). No dependence of
$\Gamma_m$ on the total depth $L_z$ suggests that the relevant
parameters in the fluidization process are just the distance from
the free surface, $z$, and $\Gamma_m$. According to this picture,
when the maximum acceleration is larger than the gravitational
acceleration ($\Gamma>1$), there is a liquid-solid interface in the
granular system. The upper part of the system is liquidlike and the
lower part is solidlike, and the position of the interface depends
on $\Gamma$.

\subsection{Relaxation near the transition point}

\begin{figure}
\center{\includegraphics[width=7.5cm]{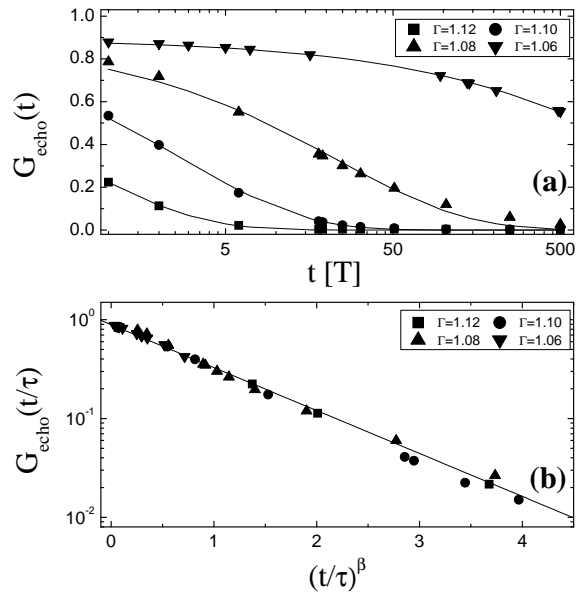}} \caption{ (a) The
height of the echoes in the correlation function up to the time of
several hundred vibration periods. Here, the position is $x=0$mm,
$z=50$mm and $f=50$Hz. (b) The same data in (a) with the scaled
horizontal axis of $(t/\tau)^{\beta}$ and the log scaled vertical
axis. \label{fig:echo}}
\end{figure}

At $1 < \Gamma < \Gamma_m $, we observe a dramatic slowing down of
the relaxation in the height of the echoes with decreasing $\Gamma$.
Figure 4(a) shows the height of the echoes of the correlation
function $G_{echo}(t)$ up to a delay time of several hundred
vibration periods. The unit of time in this figure is the vibration
period $T$. This shows that the dynamics of granular particles
undergoes structural relaxation below $\Gamma_m$. The data points
are fitted with a stretched exponential decay function,
\begin{equation}
G_{echo}(t) = G_o\exp[-(t/\tau)^{\beta}].
\end{equation}
Here, the characteristic relaxation time $\tau$ depends on $\Gamma$.
The fitting parameters of the stretched exponential decay function
are $G_o$, $\tau$, and $\beta$. During the fitting process, the
value of $G_o$ is fixed at $G_o=0.89$, which is chosen from the data
at $\Gamma=1.06$ when the delay time approaches to $0$.  In the
theory of the glass transition, the value of $G_o$ depends on the
scattering geometry and depends little on temperature and
density\cite{MCT}. The fixed value of $G_o$ is the major reason for
the poor fitting at $\Gamma=1.08$. Figure 4(b) shows the same data
with the scaled horizontal axis of $(t/\tau)^{\beta}$, where
$\beta=0.50\pm0.07$. All the data collapse well on a straight line
with a small value of the parameter $\beta$ in Eq.~(3) (Here, $\beta
=1$ implies a simple exponential relaxation\cite{Log, Glass}). This
collapse is analogous to the behavior of supercooled liquids close
to the glass transition.

\begin{figure}
\center{\includegraphics[width=7.5cm]{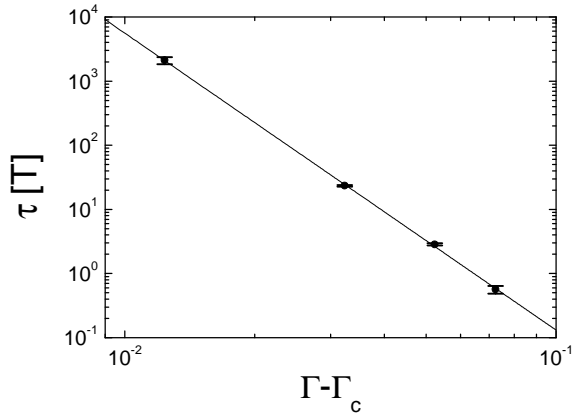}} \caption{ The
characteristic relaxation time $\tau$ of the height of the  echoes
from the same data in Fig. 4. \label{fig:time}}
\end{figure}

Figure 5 shows $\tau$ at various $\Gamma$ from the same data in
Fig.4. Here, the characteristic relaxation time $\tau$ increases
rapidly as $\Gamma$ approaches a specific value. The solid line in
this figure is a fit with the form of $\tau \sim
(\Gamma-\Gamma_c)^{-\gamma}$, where $\Gamma_c=1.048\pm0.001$ and
$\gamma=4.63\pm0.06$. The power law divergence of the characteristic
relaxation time is similar with that of mode coupling theory (MCT).
According to MCT, the characteristic relaxation time $\tau$ at a
temperature $T$ in the glassy relaxation process is of the form
$\tau \sim (T-T_c)^{-\gamma}$\cite{MCT}. Here, $T_{c}$ is a
dynamical crossover temperature from a liquidlike to a solidlike
regime, and is located between the glass transition temperature
$T_{g}$ and the melting transition temperature
$T_{m}$\cite{Nicodemi,MCT}. From the analogy between the dynamics of
the granular materials under vertical vibration and that of
supercooled liquids close to the glass transition, one expects that
the effective temperature of the granular system under vertical
vibration is related to the dimensionless acceleration. Therefore we
conjecture that $\Gamma_m$, $\Gamma_c$, and $\Gamma=1$ in the
dynamics of the granular materials under vertical vibration
correspond to $T_{m}$, $T_{c}$, and $T_{g}$ in the behavior of
supercooled liquids close to the glass transition.

\section{Conclusion}
The jamming process of a highly dense three-dimensional granular
system under vertical vibration is experimentally studied using the
DWS technique. At $\Gamma > \Gamma_m $, the granular system behaves
like fluids. At $\Gamma < 1 $, the system is arrested in the glassy
state, where the viscosity of the system appears to diverge. At $1 <
\Gamma < \Gamma_m $, the structural relaxation shows a stretched
exponential behavior and the divergence of the characteristic
relaxation time $\tau$ can be described by the power law function
which is similar to MCT. This behavior provides the evidence of the
analogy between the dynamics of granular materials and the behavior
of supercooled liquids close to the glass transition.

\begin{acknowledgments}
We would like to thank Y. H. Hwang, B. Kim, J. Lee, J. A. Seo, F.
Shan, K. W. To, and W. Goldburg for valuable discussions. This work
was supported by Grant No. R01-2002-000-00038-0 from the Basic
Research Program of the Korean Science $\&$ Engineering Foundation,
by Grant No. KRF2004-005-C00065 from the Korea Research Foundation,
and by Pusan National University in the program, Post-Doc. 2005.

\end{acknowledgments}

\end{document}